\documentclass[final,5p,times,twocolumn]{elsarticle}

\usepackage{amssymb}
\usepackage{epstopdf}
\usepackage{lineno}

\journal{ICARUS}

\makeatletter
\def\ps@pprintTitle{%
	\let\@oddhead\@empty
	\let\@evenhead\@empty
	\def\@oddfoot{}%
	\let\@evenfoot\@oddfoot}
\makeatother

\begin{document}

\begin{frontmatter}



%\title{An improved analytical model for melting probe design analysis and its application to study the velocity-dependence on gravity and the critical refreezing length}
\title{Melting probe technology for subsurface exploration of extraterrestrial ice -- Critical refreezing length and the role of gravity}


\author{K.~Sch\"uller}
\author{J.~Kowalski\corref{cor}}
\cortext[cor]{Corresponding author}
\ead{kowalski@aices.rwth-aachen.de}

\address{AICES Graduate School, RWTH Aachen University, Schinkelstr. 2, 52062 Aachen, Germany.}

\begin{abstract}

The 'Ocean Worlds' of our Solar System are covered with ice, hence the water is not directly accessible. Using melting probe technology is one of the promising technological approaches to reach those scientifically interesting water reservoirs.
Melting probes basically consist of a heated melting head on top of an elongated body that contains the scientific payload.
The traditional engineering approach to design such melting probes starts from a global energy balance around the melting head and quantifies the power necessary to sustain a specific melting velocity while preventing the probe from refreezing and stall in the channel. 
Though this approach is sufficient to design simple melting probes for terrestrial applications, it is too simplistic to study the probe's performance for environmental conditions found on some of the Ocean's Worlds, e.g. a lower value of the gravitational acceleration. This will be important, however, when designing exploration technologies for extraterrestrial purposes.

%A first order correction takes additional power losses into account, e.g. the required heat to prevent the melt channel from refreezing at the probe's hull.
%The existing analytical models, however, assume a direct contact between the melting head and the phase interface. In reality, a thin melt film evolves between the melting head and the phase interface, which causes convective losses.

We tackle the problem by explicitly modeling the physical processes in the thin melt film between the probe and the underlying ice. Our model allows to study melting regimes on bodies of different gravitational acceleration, and we explicitly compare melting regimes on Europa, Enceladus and Mars. In addition to that, our model allows to quantify the heat losses due to convective transport around the melting probe. We discuss to which extent these heat losses can be utilized to avoid the necessity of a side wall heating system to prevent stall, and introduce the notion of the 'Critical Refreezing Length'.
Our results allow to draw important conclusions towards the design of melting probe technology for future missions to icy bodies in our Solar System.
\end{abstract}


\end{frontmatter}

\section*{Nomenclature}
\noindent
\begin{tabular}{ll}
	$A$ & cross-sectional area\\
	$c_p$ & heat capacity\\
	$F$ & exerted force\\
	$F^*$ & buoyancy corrected exerted force\\
	$g$ & gravitational acceleration\\
	$h_m$ & latent heat of melting\\
	$h_m^*$ & reduced latent heat of melting\\
	$J_0,\;Y_0$ & Bessel functions of first and second kind\\
	$k$ & thermal conductivity\\
	$L$ & length of the melting probe\\
	$L^*$ & critical refreezing length\\
	$m$ & mass\\
	$n$, $d$ & fit constants\\
	$P$ & Power\\
	$p$ & pressure\\
	$p_{atm}$ & ambient pressure\\
	$\dot{Q}$ & heat flow rate\\
	$\dot{Q}_{min}$ & minimum heat flow rate\\
	$\dot{q}$ & heat flux \\
	$R$ & radius of the melting head\\
	$T$ & temperature\\
	$T_m$ & melting temperature\\
\end{tabular}

\noindent
\begin{tabular}{ll}
	$W$ & melting velocity\\
	$(r,z)$ & coordinates, see figure \ref{fig:meltingprobeschematic}\\
	$(u,w)$ & velocity components in the melt film \\
	$\textrm{Ste}$ & Stefan number\\
	$\textrm{Re}$ & Reynolds number\\
	$\alpha=k/(\rho\, c_p)$ & thermal diffusivity\\
	$\Gamma$ & boundary\\
	$\gamma$ & parameter introduced in equation (\ref{eq:gammaEquation}) \\
	$\delta$ & melt film thickness\\
	$\epsilon$ & efficiency\\
	$\eta$ & power conversion efficiency\\
	$\mu$ & dynamic viscosity of the water\\
	$\rho$ & density\\
\end{tabular}
\subsection*{Indices}
\noindent
\begin{tabular}{ll}
	$L$ & liquid phase ($\rho_L$, $c_{p,L}$, $k_L$, $\alpha_L$)\\
	$S$ & solid phase ($\rho_S$, $c_{p,S}$, $k_S$, $\alpha_S$, $T_S$)\\
	$C$ & at the phase interface ($\Gamma_C$, $\dot{Q}_C$)\\
	$H$ & at the melting head ($\Gamma_H$, $\dot{Q}_H$)\\
	$E$ & at the outflow boundary ($\Gamma_E$, $\dot{Q}_E$)
\end{tabular}
\section{Introduction}
The presence of subglacial liquid water on the icy moons of our Solar System \citep{lunine2017ocean} implies the possibility of habitable environmental conditions.
Especially the cryovolcanically active Saturnian moon Enceladus seems to be a promising candidate \citep{lunine2015enceladus} and there is some hope in the scientific community that an exploration mission to Enceladus might unravel the existence of extraterrestrial life.
Next generation mission concepts focus on orbiting and sample-returning of plume material \citep{lunine2015enceladus,sherwood2016strategic}.
Should these further strengthen any evidence for life, then the natural next step is to sample and analyze the subglacial ocean directly \citep{sherwood2016strategic,konstantinidis2015lander}.
In order to access the extraterrestrial subglacial water reservoirs, a thick ice layer must be penetrated.

A very promising technological approach for this task is to use a thermal melting probe \citep{konstantinidis2015lander}. Melting probes enforce ice penetration by heating, such that the ice in the vicinity of the probe melts and the probe eventually sinks down.
Because the amount of necessary power roughly scales with the cross-sectional area of the melting channel, a melting probe typically looks like an elongated cylinder with a heated melting head.
In comparison to other ice penetration technologies, e.g. hot water or mechanical ice drilling, the advantage of a melting probe for space exploration purposes is its smaller, lighter, and mechanically less complex design.
Melting probes are not a novel technology as they have already been applied successfully for terrestrial research since the 1960's \citep{kasser1960leichter,philberth1962geophysique}.
In recent years, however, more advanced melting probe designs have been proposed and tested \citep{zimmerman2001cryobot,stone2014progress,kowalski2016navigation,winebrenner2016clean,komle2017melting}.

A very common and relevant engineering approach to design melting probes also dates back to the 1960's and considers a straight forward energy balance \citep{aamot1967heat}:
Knowing the power $P$ implemented in a melting probe's head as well as its conversion efficiency $\eta$ allows to infer on the heat flow rate at the melting head's surface $\dot{Q} = \eta P$. The corresponding heat flux is given by $\dot{q} = \dot{Q}/A$, in which $A$ stands for the cross-sectional area of the melting head. 
The minimum heat flow rate $\dot{Q}_{min}$ required to operate the melting probe at a target melting velocity $W$ is then given by the sum of the energy necessary to increase the temperature of the ice in front of the probe and the energy that is eventually needed to melt the ice:
\begin{equation}
\label{eq:simpleEnergyBalance}
\dot{Q}_{min}=W A \rho_S\left[ h_m + c_{p,S}\left( T_m-T_S \right) \right].
\end{equation}
Here, $\rho_S$ is the ice density, $h_m$ is the latent heat of melting, $c_{p,S}$ is the heat capacity of the ice, $T_S$ is the ice temperature and $T_m$ is the melting temperature of ice. 
Note, that from equation (\ref{eq:simpleEnergyBalance}) it is now evident that the melting velocity scales inversely with the cross-sectional area of the melting probe.
In order to accommodate some scientific payload it is hence beneficial to increase the probe's length $L$ rather than its radius $R$ and often a probe design is characterized by $R<<L$.
Such an elongated geometry, however, poses another problem, namely the risk of stall due to refreezing \citep{treffer2006preliminary}.
Various concepts have been proposed to avoid stall during melting, e.g. by overheating the melting head beyond the minimally required power $\dot{Q}_{min}$ given by (\ref{eq:simpleEnergyBalance}) as proposed in \citep{aamot1967heat}, or by implementing a side wall heating system, such as realized in \citep{kowalski2016navigation}.

The amount of heat required to avoid stall, in the following referred to as the lateral heat requirement $\dot{Q}_L$, has been determined in \citep{aamot1967heat} based on quantifying heat conduction in an infinite region bounded internally by a circular cylinder \citep{jaeger1956conduction}. It reads
\begin{equation}
\label{eq:AamotLateralHeatFlowRate}
\dot{Q}_L=\frac{8 k_S T_S}{\pi}\int_0^L\int_0^\infty\frac{\exp\left( -\frac{\alpha_S b^2 z}{W} \right)}{b\left( J_0^2( R b ) Y_0^2(R b) \right)}dbdz,
\end{equation}
in which $b$ is the integration argument of the Bessel functions $J_0$ and $Y_0$, and $z$ is the spatial coordinate along the longintudinal axis of the melting probe. $L$ and
$R$ are the radius and the length of the cylindrical melting probe and $\alpha_S=k_S/(\rho_S\,c_{p,S})$ is the thermal diffusivity of the ice, in which $k_S$ denotes its thermal conductivity.

Summing up the minimum heat flow rate to open the channel $\dot{Q}_{min}$ and the lateral heat requirement $\dot{Q}_L$ provides a good approximation for the overall power necessary to sustain a specific melting velocity while preventing the probe from refreezing and stall in the channel. This 'simple' approach has been used to design thermal melting probe robots both for terrestrial field tests, e.g. in Antarctica and Greenland \citep{kowalski2016navigation,aamot1968self}, and for conceptual studies to prepare extraterrestrial exploration missions \citep{konstantinidis2015lander,zimmerman2001cryobot}.

The main technological issues for the latter have been summarized in \citep{ulamec2007access}.
In that article, the authors find that a major challenge for the design of melting probe technology for extraterrestrial purposes, is the very low ice temperatures at the target bodies, which result in a very low efficiency of the melting process.
Power efficiency is hence of major concern, especially, when facing restrictive power constraints during space missions. Efficiency can be defined as $\epsilon=\dot{Q}_{min}/(\dot{Q}_{min}+\sum_i\dot{Q}_{loss,i})$, in which $\sum_i\dot{Q}_{loss,i}$ denotes the sum of all losses. One potential loss is for example given by convective losses within the micro-scale melt film between the melting probe and the ice. The efficiency associated with these losses can either be studied experimentally \citep{KOMLE2017} or through advanced modeling techniques \citep{SCHULLER20171276} that go beyond the engineering design approach covered by (\ref{eq:simpleEnergyBalance}).

Another great challenge is associated with initiating a melting mission in a pressure regime that is below the triple point of water ($<$6.1\,mbar), in which ice sublimates if heated. This complicates the initial penetration phase of a melting probe, which is operated on bodies like Enceladus or Europa \citep{KOMLE2017}.
%Two of them are the velocity-dependence on gravity and the prevention of blocking.
After the probe reaches a certain depth, however, the melting channel is believed to refreeze and consequently the channel will sustain a pressure above the triple point, such that the probe operates in a pure melting regime \citep{treffer2006preliminary}. The initial low pressure regime can be further shortened by using a top cap as proposed in \citep{horne2017thermal}. Although a reliable and robust technological solution for the initial low pressure phase is unarguably key to a successful mission, still the larger part of the melting transit through hundreds of meters of ice will take place in a pressure regime above the triple point. Consequently, the vast amount of energy is spent in a regime that is characterized by melting rather than sublimation. This motivates to further study efficiency and dynamics of melting probe technology in a pressure regime above the triple point, while neglecting low pressure effects for the time being. 

One aspect, which has not been investigated so far even for conditions above the triple point, is the effect of the gravity on the melting process. The value of the gravitational acceleration is for example much smaller on Enceladus than it is on Europa, or even on Earth. A melting probe of a certain mass will hence exert much less force on the ice on Enceladus than it does on Europa, or on Earth. In this article, we will investigate this question by developing a physics-based model that explicitly accounts for the processes in the thin melt film between the probe and the underlying ice. This kind of theory is commonly referred to as close-contact melting theory \citep{bejan1994contact}. In section \ref{physmodel}, we will describe the physical situation for a cylindrical melting probe that heats at constant power. In section \ref{mathmodel}, we derive a model, which quantifies the conductive losses in the melt film as a function of the exerted contact force, which in itself depends on the gravitational acceleration. In contrast to the majority of other models that have been proposed for the design of melting probes, the additional consideration of exerted forces is the particular strength of our approach, which makes an investigation of the role of gravitational acceleration on the melting process possible for the first time. Section \ref{results} is devoted to results, while discussing concrete melting scenarios.
Specifically, we will apply the derived model to analyze melting regimes on Enceladus, Europa and Mars. A direct quantification of the conductive losses allows us to discuss the necessity of wall heaters to prevent stall under different extraterrestrial environmental conditions. Finally, we are able to approximate the temperature at the surface of the melting head. We will conclude with a discussion of our findings, and an outlook.
\section{Physical model}
\label{physmodel}
We consider a cylindrical melting probe of length $L$ and radius $R$ that melts through ideal ice at a constant melting velocity $W$, hence in an equilibrium melting regime.
With ideal ice we refer to polycrystalline ice without pores or fractures. Although these artifacts may be present in reality, especially close to the surface of icy moons \citep{lorenz2012thermal}, considering pure ice yields a conservative approximation for the necessary power of a melting probe.

\begin{figure}
	\centering
	\includegraphics[width=0.7\linewidth]{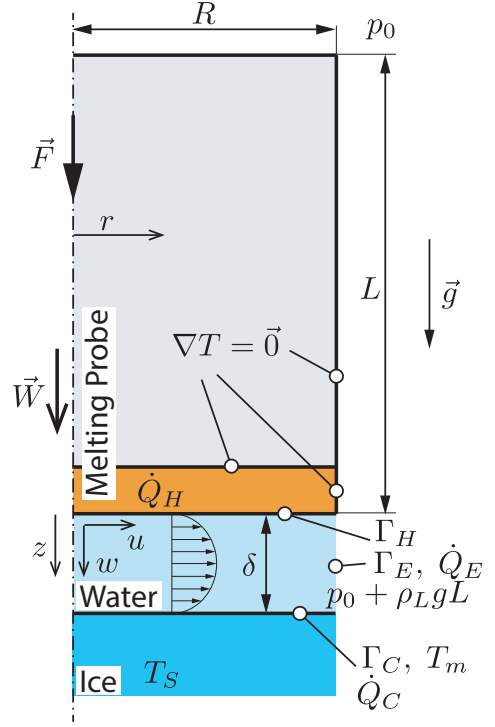}
	\caption{Schematic of the physical problem. The figure shows a cylindrical melting probe, which melts downwards. Note that the melt film thickness is not to scale.}
	\label{fig:meltingprobeschematic}
\end{figure}

The quasi-steady melting process is sketched in figure \ref{fig:meltingprobeschematic}.
A thin melt film of uniform thickness $\delta$ separates the melting head surface $\Gamma_H$, from the ice domain. The water ice interface is denoted by $\Gamma_C$. 
At $\Gamma_H$, heat is transferred from the melting head into the melt film according to $\dot{Q}_H$.
At the phase interface $\Gamma_C$, heat is transferred from the melt film into the ice according to $\dot{Q}_C$.
At the lateral outflow boundary of the melt film $\Gamma_E$, we observe convective transport of heat out of the melt film domain according to $\dot{Q}_E$.
Therefore, $\dot{Q}_E$ denotes the convective losses in the melt film, namely heat that has been transferred into the melt film but that has not been utilized to increase the probe's melting velocity.
Considering the melt film as a closed system, then yields the following energy balance
\begin{equation}
\label{eq:globalEnergyBalance}
\dot{Q}_H-\dot{Q}_E-\dot{Q}_C=0
\end{equation}
The coordinate system with axes $r$ and $z$ is fixed with respect to the melting probe.
Its origin is located at the center of the melting head surface $\Gamma_H$.
Gravitational acceleration $g$ acts in positive $z$-direction. It induces a hydrostatic pressure gradient along the probe's longitudinal axis.
A constant force $F$ is acting on the melting probe. In the absence of additional actuator systems, it would simplify to the weight of the probe that acts in $z$ direction. The exerted force causes a squeezing of the melt film, and consequently, induces a flow field within the melt film with velocity components $u$ and $w$ in $r$- and $z$-direction, respectively.

\section{Mathematical model}
\label{mathmodel}
In this work, we extend straight forward energy balance arguments to the entire melt film, i.e. to the domain bounded by $\Gamma_H$, $\Gamma_C$ and $\Gamma_E$.
%in which $\dot{Q}_H$ is the heat transport from the melting head into the melt film, $\dot{Q}_E$ is the heat flow rate that leaves the melt film at the outflow boundary and $\dot{Q}_C$ is the heat flow rate that is required for the phase change at the phase interface.
%
In order to describe the flow within the melt film, we use an extended lubrication theory \citep{moallemi1986experimental,hamrock2004fundamentals}.
The lubrication approximation is a valid simplification for the physical problem, because the melt film is very thin compared to the width of the melting head, i.e. $\delta/R<<1$.
Scaling analysis then shows that inertia terms are negligible compared to the pressure gradient, and also $\partial/\partial r^2 << \partial/\partial z^2$ as long as $\delta/R<<1$ and $(\delta/R)\textrm{Re}<<1$, in which $\textrm{Re}$ is the Reynolds number.

With these assumptions, conservation of mass, momentum and energy for the melt film are given by \citep{moallemi1986experimental}
\begin{eqnarray}
\label{eq:consOfMass}
\frac{1}{r}\frac{\partial (r\,u)}{\partial r}+\frac{\partial w}{\partial z}&=&0\\
\label{eq:consOfMomentum}
\mu\frac{\partial^2 u}{\partial z^2}&=&\frac{d p}{dr}\\
\label{eq:consOfEnergy}
u\frac{\partial T}{\partial r}+w\frac{\partial T}{\partial z}&=&\alpha_L\frac{\partial^2 T}{\partial z^2}
\end{eqnarray}
in which $p$ is the pressure, $\mu$ is the dynamic viscosity and $\alpha_L=k_L/(\rho_L\,c_{p,L})$ is the thermal diffusivity of water.

At the melting head surface $\Gamma_H$ we have no-slip/zero-inflow conditions and a prescribed heat flow rate $\dot{Q}_H$, which yields
\begin{eqnarray}
\label{eq:BCu1}
u(r,0)&=&w(r,0)=0\\
\label{eq:T0boundaryCond}
\left.\frac{\partial T}{\partial z}\right|_{z=0}&=&-\frac{\dot{Q}_H}{\pi R^2 k_L}
\end{eqnarray}
At the phase interface $\Gamma_C$, we again have no-slip in horizontal direction, however this time,
the inflow velocity is equal to the melting velocity of the probe $W$ scaled by a factor $\rho_S/\rho_L$ that accounts for the density change across the phase interface:
\begin{eqnarray}
\label{eq:BCu2}
u(r,\delta)&=&0,\qquad w(r,\delta)=-\frac{\rho_S}{\rho_L}W
\end{eqnarray}
Furthermore, two temperature related boundary conditions hold at the phase interface:
\begin{eqnarray}
\label{eq:TDeltaboundaryCond}
T(r,\delta)&=&T_m\\
\label{eq:stefanCondition}
\left.\frac{\partial T}{\partial z}\right|_{z=\delta}&=&-\frac{\rho_S W}{k_L}h_m^*
\end{eqnarray}
Here, $T_m$ is the melting temperature and equation (\ref{eq:stefanCondition}) is the so-called Stefan condition that accounts for local energy balance at the phase interface. It contains the reduced latent heat of melting $h_m^*= h_m+c_{p,S}(T_m-T_S)$, which is the sum of the latent heat $h_m$ and the specific energy necessary to increase the temperature of the ice. 

For a vertically oriented melting probe in a water filled melting channel, the pressure increases linearly along the probe's length, which causes the pressure at the outflow boundary ($r=R$) to be
\begin{equation}
\label{eq:atmPressure}
p(R)=p_0+\rho_L g L
\end{equation}
in which $p_{0}$ is the pressure at the back of the probe.

\subsection{Heat flow rate at the phase interface $\dot{Q}_C$}
At the phase interface $\Gamma_C$, the heat flow rate equals the heat flux multiplied by the probe's cross-sectional area
\begin{equation}
\label{eq:phaseInterfaceHeatFlowRate}
\dot{Q}_C=-\pi R^2 k_L\left.\frac{\partial T}{\partial z}\right|_{z=\delta}
\end{equation}
Together with the Stefan condition (\ref{eq:stefanCondition}) this yields
\begin{equation}
\label{eq:phaseInterfaceHeatFlowRateFinal}
\dot{Q}_C=\pi R^2 \rho_S W h_m^*
\end{equation}
Hence, equation (\ref{eq:phaseInterfaceHeatFlowRateFinal}) is formally equivalent to the straight-forward energy balance (\ref{eq:simpleEnergyBalance}) discussed in the introduction. The important innovation in this paper is that we consider the realistic case, in which less power is utilized for forward motion than is inputted into the system, hence $\dot{Q}_C < \dot{Q}_H$. The next section will be devoted to quantifying the heat losses $\dot{Q}_E = \dot{Q}_H - \dot{Q}_C$.

\subsection{Heat flow rate at the outflow boundary $\dot{Q}_E$}
To avoid unnecessary complications, we will assume that all thermo-physical properties are phase-wise constant. Integrating the momentum balance (\ref{eq:consOfMomentum}) twice with respect to $z$ and substituting in the boundary conditions~(\ref{eq:BCu1}) and~(\ref{eq:BCu2}) yields the horizontal melt film velocity
\begin{equation}
\label{eq:horizontalVeloWithPressureGradient}
u=\frac{1}{2\mu}\frac{dp}{dr} z (z-\delta).
\end{equation}
Substituting the velocity (\ref{eq:horizontalVeloWithPressureGradient}) into continuity equation (\ref{eq:consOfMass}), integrating the resulting equation over the melt film thickness and inserting the boundary conditions (\ref{eq:BCu1}) and (\ref{eq:BCu2}) yields
\begin{equation}
	\label{pressureGradient0}
	-\frac{\rho_S}{\rho_L} W\int_0^r \tilde{r} d\tilde{r}=\frac{r}{12 \mu}\frac{dp}{dr}\delta^3.
\end{equation}
The pressure gradient is found by integrating equation (\ref{pressureGradient0}) and making use of $dp/dr=0$ at $r=0$ due to axisymmetry:
\begin{equation}
\label{eq:pressureGradientFinal}
\frac{dp}{dr}=-\frac{6\mu\frac{\rho_S}{\rho_L}W r}{\delta^3}
\end{equation}
The horizontal velocity component (\ref{eq:horizontalVeloWithPressureGradient}) is then evaluated to be
\begin{equation}
\label{eq:horizontalVeloFinal}
u=-\frac{3\frac{\rho_S}{\rho_L}W r z(z-\delta)}{\delta^3}.
\end{equation}
To derive an equation for the melt film thickness, we at first use continuity to write the temperature equation in conservation form.
%add the continuity equation (\ref{eq:consOfMass}), multiplied with the temperature $T$, to the left hand side of the conservation of energy (\ref{eq:consOfEnergy}), which yields
%\begin{equation}
%\label{eq:consOfEnergy2}
%u\frac{\partial T}{\partial r}+w\frac{\partial T}{\partial z}+T\left( \frac{1}{r}\frac{\partial \left( r\,u \right)}{\partial r}+\frac{\partial w}{\partial z} \right)=\alpha_L\frac{\partial^2 T}{\partial z^2}
%\end{equation}
%Equation (\ref{eq:consOfEnergy2}) can be rewritten by applying product rule. 
Integration over the melt film thickness then yields
\begin{equation}
\label{eq:consOfEnergy3}
\int_{0}^{\delta}\left[\frac{1}{r}\frac{\partial \left(r\,u\,T\right)}{\partial r}+\frac{\partial \left(w\,T\right)}{\partial z}\right]dz=\alpha_L\int_{0}^{\delta}\frac{\partial^2 T}{\partial z^2}dz
\end{equation}
This can be simplified by substituting in boundary conditions (\ref{eq:BCu1})--(\ref{eq:stefanCondition}):
\begin{eqnarray}
\label{eq:consOfEnergy4}
\frac{d}{rd r}\int_{0}^{\delta} r\,u\,T dz -\frac{\rho_S}{\rho_L}W T_m=-\alpha_L\left( \frac{\rho_SW}{k_L}h_m^*+\left. \frac{\partial T}{\partial z} \right|_{z=0} \right)
\end{eqnarray}
We make a quadratic polynomial Ansatz for the temperature field in $z$-direction, which satisfies the three boundary conditions (\ref{eq:T0boundaryCond}), (\ref{eq:TDeltaboundaryCond}) and (\ref{eq:stefanCondition}). The resulting polynomial is
\begin{eqnarray}
\label{eq:quadraticPolynomialTemperature}
T=\frac{\dot{q}_H-\rho_S W h_m^*}{2 \delta k_L} \left(z^2 -\delta^2\right)+\frac{\dot{q}_H}{k_L}\left( \delta-z \right)+T_m
\end{eqnarray}
in which we recall that
\begin{equation}
\label{eq:heatFluxDefinition}
\dot{q}_H=\dot{Q}_H/(\pi R^2)
\end{equation}
is the heat flux at the melting head.
Substituting the temperature Ansatz (\ref{eq:quadraticPolynomialTemperature}) and the horizontal velocity (\ref{eq:horizontalVeloFinal}) into the conservation of energy (\ref{eq:consOfEnergy4}), finally yields an expression for the melt film thickness $\delta$
\begin{eqnarray}
\label{eq:meltFilmThicknessQuadratic}
%\tilde{\delta}^2+\tilde{\delta}\frac{3\,\mathrm{Ste}+20}{2\tilde{W}}-\frac{10\,\mathrm{Ste}}{\tilde{W}^2}=0
\delta=\frac{20\alpha_L \rho_L \left( \dot{q}_H-\rho_S W h_m^* \right)}{W \rho_S \left( 3 \dot{q}_H+7 \rho_S W h_m^* \right)}
\end{eqnarray}
%Substituting the heat flow rate at the phase interface (\ref{eq:phaseInterfaceHeatFlowRateFinal}) and the heat flux at the melting head (\ref{eq:heatFluxDefinition}) into equation (\ref{eq:meltFilmThicknessQuadratic}) results in
which can also be written in terms of the heat flow rates at the melting head surface and at the phase interface:
\begin{equation}
	\label{eq:deltaFSte}
	\delta=\frac{20\alpha_L \rho_L \left( \dot{Q}_H-\dot{Q}_C \right)}{W \rho_S \left( 3 \dot{Q}_H+7 \dot{Q}_C \right)}
\end{equation}
Still $\delta$ cannot be computed as the melting velocity is not known. In order to close the system, we consider the force balance as an additional relation. It states that the integral of the pressure in the melt film must balance the exerted force $F$ (e.g. given by the gravity proportional weight of the melting probe). This is
\begin{equation}
\label{eq:forceBalanceTemp}
F=\int_0^{R}\int_0^{2\pi} (p-p_{0}) r d\phi dr
\end{equation}
An expression for the pressure is given by integrating the pressure gradient (\ref{eq:pressureGradientFinal}) subject to the pressure boundary condition (\ref{eq:atmPressure}), which yields
\begin{equation}
\label{eq:pressureEquationFinal}
p-p_{0}=\frac{3\mu\frac{\rho_S}{\rho_L}W (R^2-r^2)}{\delta^3}+\rho_L g L.
\end{equation}
Substitution into the force balance (\ref{eq:forceBalanceTemp}) again results in an equation for the melt film thickness $\delta$
\begin{equation}
\label{eq:meltFilmThicknessFromForceBalance}
\delta=\left( \frac{3}{2}\frac{\pi R^4 \mu \frac{\rho_S}{\rho_L} W}{F^*} \right)^{1/3}
\end{equation}
in which $F^*=F-\pi R^2\rho_L g L$ is the buoyancy corrected force. Note, that equation (\ref{eq:meltFilmThicknessFromForceBalance}) is a mechanical approximation to the melt film thickness $\delta$, while the previous equation (\ref{eq:deltaFSte}) has been a thermodynamic approximation. Equating both yields a relation that allows to quantify the fraction of the input heat that is utilized for the forward motion of the probe  
\begin{equation}
	\label{eq:finalQEequation}
	%\dot{Q}_H=\frac{7\gamma+1}{1-3\gamma} \dot{Q}_C
	\dot{Q}_C=\frac{1-3\gamma}{7\gamma+1} \dot{Q}_H,
\end{equation}
in which
\begin{equation}
	\label{eq:gammaEquation}
	\gamma=\frac{1}{20 \alpha_L}\left( \frac{\rho_S}{\rho_L} W R \right)^{4/3} \left( \frac{3 \pi \mu}{2 F^*} \right)^{1/3}.
\end{equation}
\subsection{Model summary}
Equation (\ref{eq:finalQEequation}) finally allows us to determine the convective losses at the melt film outflow boundary $\dot{Q}_E$
\begin{equation}
\label{eq:exitFlowRate3}
	%\dot{Q}_E=\dot{Q}_C\left( \frac{7\gamma+1}{1-3\gamma}-1 \right)
	\dot{Q}_E=\dot{Q}_H-\dot{Q}_C =\left( 1- \frac{1-3\gamma}{7\gamma+1} \right)\dot{Q}_H
\end{equation}
%
%In the previous subsections, we derived equations for the heat flow rate at the phase interface $\dot{Q}_C$ (\ref{eq:phaseInterfaceHeatFlowRateFinal}) and at the outflow boundary $\dot{Q}_E$ (\ref{eq:exitFlowRate3}).
%Substituting these equations into the energy balance of the melt film (\ref{eq:globalEnergyBalance}) yields
Our results allow us to split the heat flow rate at the melting head surface into the portion of heat that is used for the forward motion of the melting probe $\dot{Q}_C$, and additional convective losses given as a fraction of $\dot{Q}_C$
\begin{equation}
%\label{eq:finalQEequation}
\label{eq:finalQHEquation}
\dot{Q}_H=\dot{Q}_C+\left( \frac{7\gamma+1}{1-3\gamma}-1 \right)\dot{Q}_C
\end{equation}
This relation can be used in two ways: One can either determine the required heat flow rate at the melting head $\dot{Q}_H$ to sustain a target melting velocity for a given exerted force. The factor 
$(1-3\gamma)/(7\gamma+1)$ is then to be interpreted as an efficiency factor less than 100\,\%, which has to be compensated. Alternatively, one can determine the effective melting velocity, given a certain input power $\dot{Q}_H$. Since both $\gamma$ and $\dot{Q}_C$ depend on the $W$, the latter requires a numerial root finding strategy.
\section{Results and discussion}
\label{results}
In this section, we apply our model to analyze the melting process of a cylindrical melting probe.
We choose a length of $L=1$\,m, a radius of $R=0.06$\,m and a mass of $m=25$\,kg.
These dimensions correspond to both the IceMole flight model \citep{konstantinidis2015lander} and the Cryobot \citep{zimmerman2001cryobot}, both of which have been designed for extraterrestrial application.
%According to the scope of our model, we assume an isothermal planar melting head and polycrystalline ice with no pores or fractures.
We consider three possible extraterrestrial targets, namely the polar caps on Mars, the south polar region on Enceladus and Europa. The relevant thermo-physical properties are summarized in table \ref{tab:designParametersAndProperties}. Within the scope of this study and as discussed in the introduction, we will focus on pure melting without sublimation.
%We would like to stress again that we only look at pure melting without sublimation, which will occur during the initial melting process due to the low environmental pressure.

\begin{table}
	\caption{Thermo-physical properties. Data from \citep{konstantinidis2015lander} and \citep{ulamec2007access}.}
	\label{tab:designParametersAndProperties}
	\centering
	\begin{tabular}{lrrr}
		\hline
		 & Mars (polar) & Enceladus & Europa \\
		\hline
		$g$ [m/s$^2$] & 3.7 & 0.1 & 1.3\\
		$k_L$ [W/(m\,K)] & 0.6 & 0.6 & 0.6\\
		$\rho_L$ [kg/m$^3$] & 1000 & 1000 & 1000\\
		$c_{p,L}$ [J/(kg\,K)] & 4200  & 4200 & 4200 \\
		$\mu$ [N\,s/m$^2$] & 0.0013 & 0.0013 & 0.0013 \\
		$k_S$ [W/(m\,K)] & 2.5 & 3.0 & 3.5 \\
		$\rho_S$ [kg/m$^3$] & 921.3 & 925.1 & 927.8\\
		$c_{p,S}$ [J/(kg\,K)] & 1877 & 1656 & 1476 \\
		$h_m$ [J/kg] & 333700 & 333700 & 333700\\
		$T_m$ [K] & 273 & 273 & 273\\
		$T_S$ [K] & 210 & 150 & 100\\
		\hline
	\end{tabular}
\end{table}

\subsection{The role of gravitational acceleration}
In the absence of an actuator system, the buoyancy corrected force of the melting probe $F^*$ is given by its weight and a buoyancy term. Both are proportional to the gravitational acceleration by definition.
Studying the role of gravitational acceleration on the melting process, hence translates into studying its force dependency. The effect of $F^*$ on the melting process can directly be observed by analyzing the melt film thickness $\delta$ (\ref{eq:meltFilmThicknessFromForceBalance}) and the heat flow rate at the outflow boundary $\dot{Q}_E$ (\ref{eq:exitFlowRate3}).
For heavy melting probes or environments that exhibit a large gravitational acceleration, the force $F^*$ increases, and the melt film thickness tends to zero. 
Consequently, convective losses $\dot{Q}_E$ get small, which means that melting gets efficient and $\dot{Q}_H \approx \dot{Q}_C$.
This corresponds to the straight-forward energy balance (\ref{eq:simpleEnergyBalance}), which is assumed to operate at optimal efficiency, and hence is a special case of our model.
If on the other hand exerted forces get small, either for very light melting probes or environments that exhibit a small gravitational acceleration, then the convective losses increase. The limiting case $F^* \rightarrow 0$ deserves further attention, as it results in the intuitive solution of a stagnant melting probe ($W\rightarrow 0$). While the melting head remains thermally active, the phase interface moves gradually away from the melting probe without any motion of probe itself. This physical situation is commonly referred to as a Stefan problem.

\begin{figure}
	\centering
	\includegraphics[width=1\linewidth]{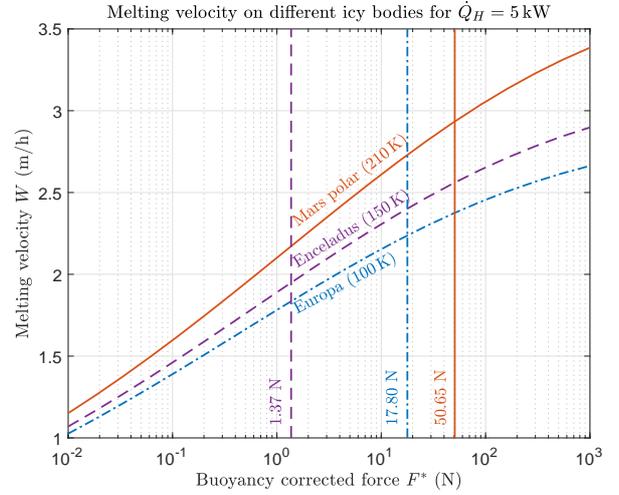}
	\caption{Results for a melting probe ($L=1\,$m, $R=0.06\,$m) with a heat flow rate at the melting head of $\dot{Q}_H=5\,$kW. The plot shows the melting velocity over the buoyancy corrected force $F^*=F-\pi R^2\rho_L g L$ on different extraterrestrial ice environments. The vertical lines mark the buoyancy corrected force at the target.}
	\label{fig:meltingvelocity5kw}
\end{figure}

\begin{figure}
	\centering
	\includegraphics[width=1\linewidth]{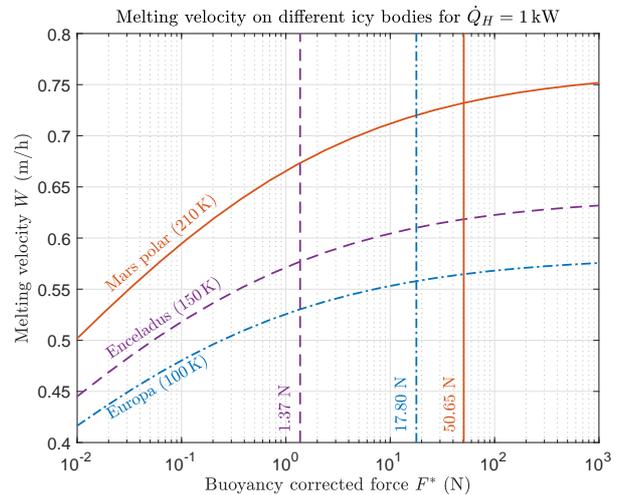}
	\caption{Results for a melting probe ($L=1\,$m, $R=0.06\,$m) with a heat flow rate at the melting head of $\dot{Q}_H=1\,$kW. The plot shows the melting velocity over the buoyancy corrected force $F^*=F-\pi R^2\rho_L g L$ on different extraterrestrial ice environments. The vertical lines mark the buoyancy corrected force at the target.}
	\label{fig:meltingvelocity1kw}
\end{figure}

Figure \ref{fig:meltingvelocity5kw} and \ref{fig:meltingvelocity1kw} show how the melting velocity varies with the buoyancy corrected force for environmental conditions found on Mars, Europa and Enceladus (see table \ref{tab:designParametersAndProperties}). Figure \ref{fig:meltingvelocity5kw} assumes a heat flow rate at the melting head of $\dot{Q}_H=5\,$kW, while figure \ref{fig:meltingvelocity1kw} assumes $\dot{Q}_H=1\,$kW. The vertical lines mark the buoyancy corrected forces for a melting probe of 25kg, which are $50.65\,$N on Mars, $1.37\,$N on Enceladus and $17.8\,$N on Europa.
Equation (\ref{eq:finalQHEquation}) was solved numerically in order to obtain the melting velocities in the figures.

The plots clearly indicate that the melting velocity monotonically increases with increasing force. In the limit ($F^*\rightarrow\infty$) the melting velocity converges asymptotically to a value that corresponds to the velocity obtained through the straight-forward energy balance (\ref{eq:simpleEnergyBalance}). The latter observation is more pronounced in the $\dot{Q}_H=1\,$kW case, as the influence of reasonable force variations (1\,N -- 1000\,N) on the melting velocity is larger if the heat flow rate at the melting head is large. Increasing the force spectrum considered in the plot for the $\dot{Q}_H=5\,$kW case would however result in a similar asymptotic behavior.
An additional observation is also that the melting velocity increases with increasing ice temperature.
The highest melting velocity for each force value is consistently found for Mars environmental conditions that are the warmest, the lowest for Europa conditions, which is also the coldest of the three.
An interesting finding of our model is the combined impact of ice temperature and gravitational acceleration. Despite of melting through much colder ice on Europa, our 25\,kg probe still operates at a higher velocity than on Enceladus, which is warmer but has a significantly lower gravitational acceleration. Hence, convective losses due to a low value of the exerted contact force seem to dominate the effect of the warmer ice on Enceladus.
Our results refute earlier findings reported in \citep{ulamec2007access} and based on \citep{shreve1962theory} that the force of gravity is necessary to set the direction of penetration, but does not (to first order) control the melting velocity. Increasing the melting velocity of a probe on Enceladus hence cannot be achieved by increasing its mass alone, as the high masses needed would not be feasible for a space mission that starts on Earth. One solution would be to introduce an additional force by means of an actuator system, e.g. an ice screw at the melting head, as it is realized with the IceMole melting probe \citep{kowalski2016navigation}.

\subsection{Critical refreezing length}
One key question during the design of melting probes is whether a side wall heating system must be implemented in order to prevent refreezing at the probe's hull. A wrong design could cause stall of the melting probe and poses a severe risk to a melting probe mission. To mitigate this risk, a common approach is to over design the melting probe by utilizing wall heaters, even if they would not be needed. 

In this section, we will investigate to which extent the convective melt film losses naturally contribute to prevent refreezing. Our analysis is based on an approximation to the lateral heat requirement (\ref{eq:AamotLateralHeatFlowRate}) derived by \citep{aamot1967heat,ulamec2007access}
\begin{eqnarray}
\label{eq:lateralHeatRequirementApprox}
\frac{\dot{Q}_L}{R^2 W (T_m-T_S)}&=&n\,\sigma^d\quad\textrm{with}\quad \sigma=\frac{L^*}{W R^2}
\end{eqnarray}
with the fit constants $n=932$\,Ws/K/m$^3$ and $d = 0.726$. According to \citep{ulamec2007access}, equation (\ref{eq:lateralHeatRequirementApprox}) is valid for $5 \times 10^4\,\textrm{s/m}^2 < \sigma < 10^8\,\textrm{s/m}^2$.
Note the difference between the actual length of the probe $L$ and $L^*$ used in equation (\ref{eq:lateralHeatRequirementApprox}). $L^*$ stands for the distance along the melting channel after which the temperature has relaxed to melting temperature, hence the distance from the melting head after which refreezing would occur. It will be referred to as the critical refreezing length in the ongoing of this article.

Assuming that the convective losses of the melt film $\dot{Q}_E$ contribute towards keeping the lateral melt channel open, we can investigate $L^*$ for
\begin{equation}
	\label{eq:reqToPreventStall}
	\dot{Q}_L = \dot{Q}_E.
\end{equation}
Substituting equations (\ref{eq:exitFlowRate3}) and (\ref{eq:lateralHeatRequirementApprox}) into equation (\ref{eq:reqToPreventStall}) yields

\begin{equation}
\label{eq:criticallengthEquation}
%L\leq L^*=\left[ \frac{R^{2(d-1)}W^{d-1}\left( \dot{Q}_H-\pi R^2 \rho_S W h_m^* \right)}{n(T_m-T_S)} \right]^{1/d}
%L^*=\left[ \frac{c_{p,L} h_m^*}{n k_L (T_m-T_S)}\left( \frac{3}{16} \frac{\mu\rho_S^7\pi^4 R^{6d+4}W^{3d+4}}{\rho_L F^*} \right)^{1/3} \right]^{1/d}
L^*=\left[ \frac{\dot{Q}_C W^{d-1} R^{2\left( d-1 \right)}}{n\left( T_m-T_S \right)}\left( \frac{7\gamma+1}{1-3\gamma}-1 \right) \right]^{1/d}.
\end{equation}
Hence, wall heaters to prevent refreezing must be implemented if the probe is shorter than $L^*$.

\begin{figure}
	\centering
	\includegraphics[width=1\linewidth]{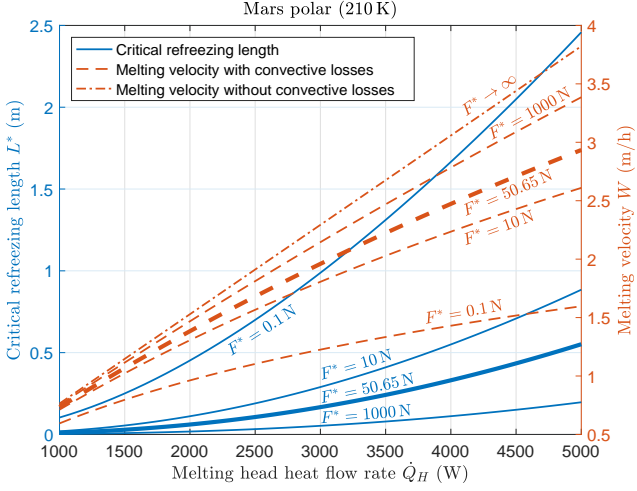}
	\caption{Critical refreezing length and melting velocity over the heat flow rate at the melting head for different buoyancy corrected forces in $210\,$K cold ice on Mars.}
	\label{fig:criticallengthmars}
\end{figure}

\begin{figure}
	\centering
	\includegraphics[width=1\linewidth]{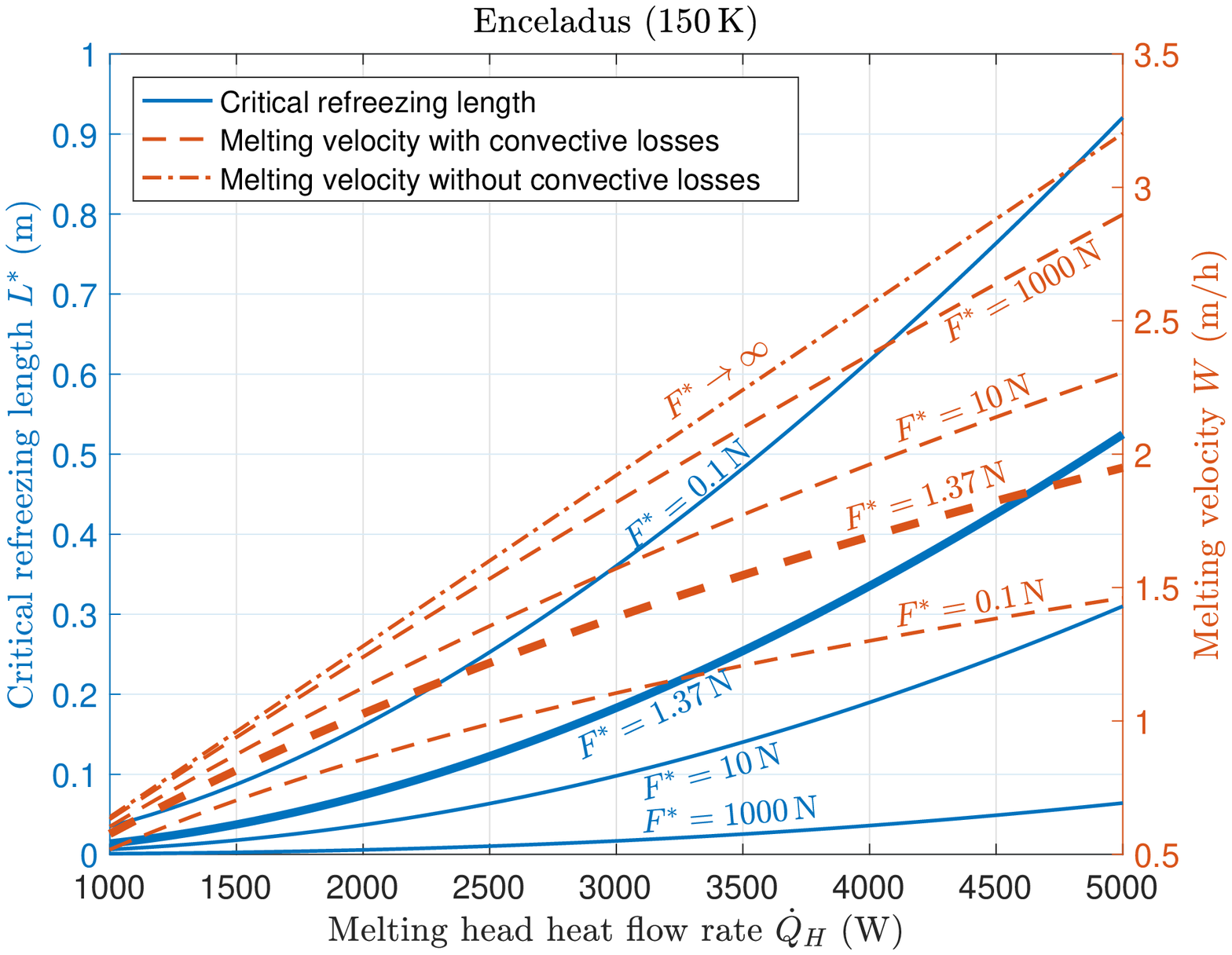}
	\caption{Critical refreezing length and melting velocity over the heat flow rate at the melting head for different buoyancy corrected forces in $150\,$K cold ice on Enceladus.}
	\label{fig:criticallengthenceladus}
\end{figure}

\begin{figure}
	\centering
	\includegraphics[width=1\linewidth]{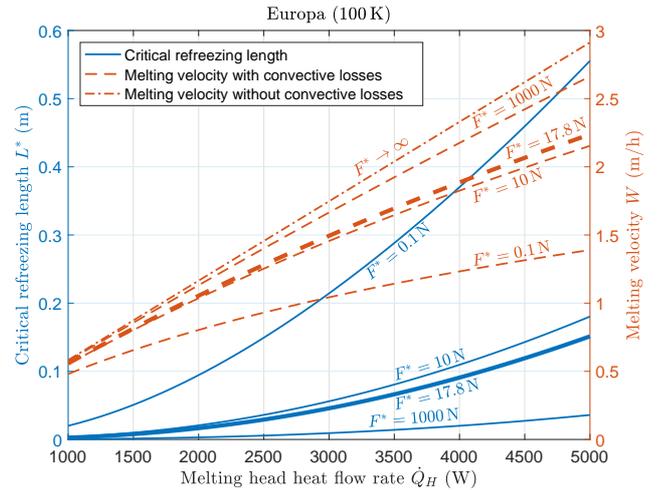}
	\caption{Critical refreezing length and melting velocity over the heat flow rate at the melting head for different buoyancy corrected forces in $100\,$K cold ice on Europa.}
	\label{fig:criticallengtheuropa}
\end{figure}

Figures \ref{fig:criticallengthmars}, \ref{fig:criticallengthenceladus} and \ref{fig:criticallengtheuropa} show the critical refreezing length for the melting probe using equation (\ref{eq:criticallengthEquation}) in combination with the melting velocity for Mars polar, Enceladus and Europa environmental conditions, respectively.
Note, that for a constant heat flow rate at the melting head, the critical refreezing length $L^*$ increases if the exerted force decreases. 
This can be explained by looking at the melting velocity, which depends on the exerted force.
When the melting probe slows down, the melt film thickness increases and higher convective losses occur. For one value of the buoyancy corrected force, the critical refreezing length increases with increasing heat flow rate at the melting head. Again, this is a result of increasing convective losses.

When comparing the results for 210\,K (figure \ref{fig:criticallengthmars}), 150\,K (figure \ref{fig:criticallengthenceladus}) and 100\,K (figure \ref{fig:criticallengtheuropa}), it can be seen that the critical refreezing length is smaller in a cold environment.
This is because more heat ($\dot{Q}_L\propto(T_m-T_S)$, see equation (\ref{eq:lateralHeatRequirementApprox})) is needed at the probe's hull to prevent stall.
Our melting probe has a length of $1\,$m.
Since the critical refreezing length is always smaller than $1$\,m for reasonable buoyancy corrected forces ($F^*>10\,$N), the melting probe will stall if no additional wall heaters are provided.

\subsection{Optimal melting probe length}
In the previous subsection, the critical refreezing length $L^*$ has been introduced as a measure to check if a melting probe of given length, radius and mass needs wall heaters, and at which minimal distance from the head these must be implemented.
If the critical refreezing length is smaller than the length of the melting probe, the natural next question is: What probe length would make the use of a side wall heating system dispensable. 
%In the case of a variable melting probe length, one could reduce its length until it equals the critical refreezing length.
Note, that this probe length is not given by the aforementioned critical refreeze length, as length of the probe itself alters the model. Instead, the critical refreezing length $L^*$ in equation (\ref{eq:criticallengthEquation}) is at replaced by the length of the probe $L$.
The resulting equation is
\begin{equation}
\label{eq:nonlinearCriticalLengthEq}
0=L-\left[ \frac{\dot{Q}_C W^{d-1} R^{2\left( d-1 \right)}}{n\left( T_m-T_S \right)}\left( \frac{7\gamma+1}{1-3\gamma}-1 \right) \right]^{1/d}
\end{equation}
Note that $\gamma$ and the melting velocity are now functions of the buoyancy corrected force $F^*$ and hence depend on the melting probe length $L$. Therefore, equation (\ref{eq:nonlinearCriticalLengthEq}) has to be solved numerically for $L$.

If the mass of a melting probe is given and no actuators are used to increase the exerted force of the melting probe, the definition of the buoyancy corrected force can be exploited to derive an upper limit for the length of the melting probe as a function of its mass and radius. Motion is only possible if the buoyancy corrected force is larger than zero, which yields the relation
\begin{equation}
	\label{eq:upperLLimit}
	L< \frac{m}{\pi R^2 \rho_L}
\end{equation}
For the melting probe of 25\,kg and $R=0.06$\,m, equation (\ref{eq:upperLLimit}) yields an upper limit for the length of approximately $2.21$\,m, if $\rho_L=1000$\,kg/m$^3$.

\begin{figure}
	\centering
	\includegraphics[width=1\linewidth]{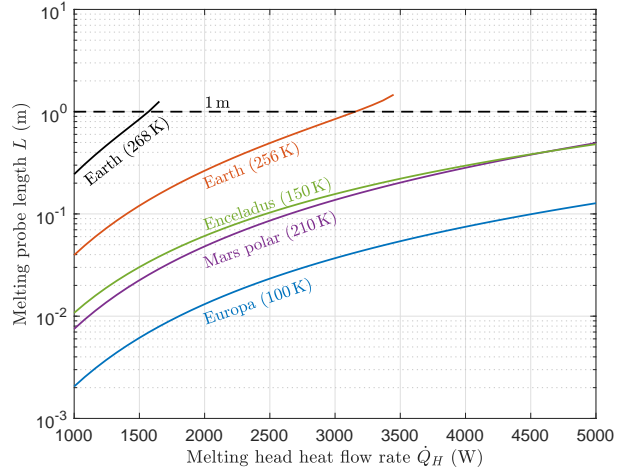}
	\caption{Maximum melting probe length for a 25\,kg and $R=0.06$\,m melting probe that operates without a side wall heating system in different ice environments. The dashed line is used to mark the $1$\,m level in the plot.}
	\label{fig:meltingprobelengths}
\end{figure}

Figure \ref{fig:meltingprobelengths} shows the maximum length of a melting probe ($m=25$\,kg, $R=0.06$\,m) that can operate without side wall heaters for different ice environments. As expected, the probe's length can be significantly larger for warm ice. Therefore, most melting probes for temperate ice on Earth (i.e. for ice with a temperature close of 273.15\,K) do not need wall heaters. However, there are also colder ices on Earth with temperatures in the order of 256\,K, e.g. in Antarctica. If the heat flow rate at the melting head is smaller than approximately 3250\,W, the melting probe needs to be smaller than 1\,m. Otherwise a wall heating system must be utilized to prevent stall of the probe. If the heat flow rate is larger than 3250\,W, the resulting heat flow rate at the outflow boundary $\dot{Q}_E$ is sufficient to prevent stall of the probe in 256\,K cold ice.

Even for the high power scenario $\dot{Q}_H$=5kW, we observe that the maximum melting probe length for all extraterrestrial environments is less than 0.5\,m. Interestingly, the melting probe length for Enceladus is very close to the one for Mars. This is again due to the combined effect of the low gravitational acceleration and the difference in ice temperature. A small gravitational acceleration, as it is present on Enceladus compared to Mars, will decrease the buoyancy corrected force and hence increase the convective losses $\dot{Q}_E$. On the other hand, the ice temperature on Enceladus is significantly less than on Mars. Therefore, more heat is required at the walls to keep the water from refreezing. Interestingly both effects cancel out for Mars and Enceladus conditions.
The maximum melting probe length for Europa is in the order of 0.1\,m for $\dot{Q}_H=5000$\,W. A probe of this dimension and a mass of 25\,kg including scientific instruments is technically not feasible. We therefore conclude that melting probes for future Europa missions must be utilized with wall heaters, while it might be possible to design a system for Enceladus and Mars without additional side wall heaters.

\subsection{Melting head temperature and Stefan number}
\label{sec:StefanNumber}

\begin{figure}
	\centering
	\includegraphics[width=1\linewidth]{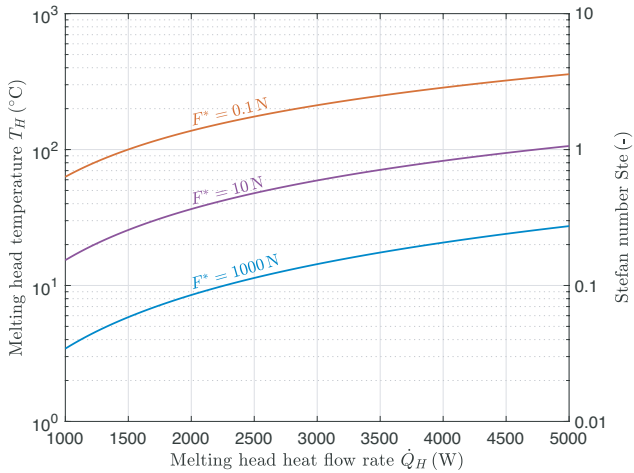}
	\caption{Melting head temperatures and Stefan numbers as a function of heat flow rate at the melting head for Enceladus conditions. The three lines correspond to three different values of the buoyancy corrected force, namely 0.1\,N, 10\,N and 1000\,N. The figure shows the results using equation (\ref{eq:StefanCondFinal}).}
	\label{fig:stefannumberplotpost}
\end{figure}

The temperature at the melting head as predicted by our model is found by evaluating the quadratic temperature profile (\ref{eq:quadraticPolynomialTemperature}) at $z=0$, while making use of the computed melt film thickness (\ref{eq:deltaFSte}). This yields

\begin{equation}
\label{eq:temperatureOfMeltingHead}
T_H=\frac{10\left( \dot{Q}^2_H -\dot{Q}^2_C \right)}{c_{p,L}\rho_S \pi R^2 W \left( 3\dot{Q}_H+7\dot{Q}_C \right)}+T_m
\end{equation}
Figure \ref{fig:stefannumberplotpost} shows the melting head temperature for three force scenarios as expected on Enceladus conditions.
Knowing the surface temperature $T_H$ also allows us to quantify the Stefan number regime for our melting process. The Stefan number is typically defined as $\textrm{Ste}=c_p (T_H-T_m)/h_m^*$ and relates the sensible heat to the reduced latent heat of the considered system. Since the task of a melting probe is to enforce phase change of the surrounding ice, sensible heat in the system is undesirable and can hence be interpreted as a measure of inefficiency. Therefore, small Stefan numbers indicate efficient melting. 
%Another feature of small Stefan numbers ($\textrm{Ste}<<1$) in terms of close-contact melting is that the temperature in the melt film reduces to a linear profile in $z-$direction \citep{emerman1983stokes}.

Based on the evaluated surface temperature of the melting head (\ref{eq:temperatureOfMeltingHead}), the Stefan number can be translated in
\begin{equation}
\label{eq:StefanCondFinal}
\textrm{Ste}=\frac{10\left( \dot{Q}_H^2-\dot{Q}_C^2 \right)}{\dot{Q}_C \left( 3\dot{Q}_H+7\dot{Q}_C \right)}
\end{equation}
In addition to the melting head surface temperatures, figure \ref{fig:stefannumberplotpost} also shows the Stefan numbers over the the heat flow rate at the melting head for three different buoyancy corrected forces. The curves have been produced using equation (\ref{eq:StefanCondFinal}) with the thermo-physical properties for water ice on Enceladus. Since the Stefan number $\textrm{Ste}$ is just a different scaling of the melting head temperature, the curves are identical.

It can be seen that the temperature at the melting head $T_H$ increases with increasing heat flow rate $\dot{Q}_H$ and with decreasing buoyancy corrected force $F^*$. For $F^*=0.1\,$N, the temperature exceeds $100\,^\circ$C at $\dot{Q}_H\approx 1500\,$W, which at an ambient pressure of $1\,$bar would induce boiling. Note, that a liquid-gaseous phase change is beyond the capabilities of the presented model and therefore this curve should be interpreted with care. The melting head temperature for $F^*=10\,$N is between $15\,^\circ$C at $\dot{Q}_H=1000\,$W and $94\,^\circ$C at $\dot{Q}_H=4500\,$W. Such high temperatures can be dangerous if the maximum survival temperature of the utilized electronic components in the melting head is exceeded. To overcome this risk, a temperature feedback control might be implemented to switch off the melting head heaters if a critical temperature is exceeded. This in return, would of course reduce the melting velocity and blow up the overall penetration time. The lowest temperatures are obtained for $F^*=1000\,$N. For this relatively high buoyancy corrected force, the temperature varies between $3\,^\circ$C at $\dot{Q}_H=1000\,$W and $28\,^\circ$C at $\dot{Q}_H=5000\,$W.

%It can be seen that $\textrm{Ste}<1$ for $F^*>10\,$N for the considered range of $\dot{Q}_H$.
\section{Conclusions}
We applied close-contact melting theory to develop an improved model for the design of melting probes that allows to study the role of gravitational acceleration as well as the melting heads surface temperature. Our model explicitly accounts for the physical processes in the melt film between the melting head and the water ice interface. It allows a direct quantification of the convective losses in the melt film, and hence provides a more accurate description of the efficiency of the melting process than a so far often used straight-forward energy balance.
%The basic idea is to use a global energy balance in the melt film coupled to conservation of mass and momentum of thin film flows.
Furthermore, we introduced the notion of the critical refreezing length and used it to discuss to which extent convective losses during the melting process can naturally be utilized to avoid otherwise over designed side wall heating systems.
The concept of the critical refreezing length is not only an important measure for melting probes, but also important, when mitigating the risk due to radioisotope power sources in the context of planetary protection on the icy bodies in our Solar System as described in \citep{lorenz2012thermal}. 
%For this application, a small critical refreezing length is desirable.
%Applying our model to radioisotope power sources of spacecrafts could provide a better estimate whether a broken spacecraft could be a risk for contamination.

The three major findings of our work are:

\begin{itemize}
	\item Other than previous literature has stated, we find that the exerted force, hence the gravitational acceleration of the environment can have a significant effect on the melting velocity. In particular this is the case, if the power input, hence the heat flow rate at the melting head is large. Small mass budgets for extraterrestrial exploration missions hence motivate the use of actuator systems, e.g. an ice screw as proposed in \citep{kowalski2016navigation} in order to superpose the probe's intrinsic weight with an additional force.
	\item Wall heaters must be considered for the design of extraterrestrial melting probes, at least if less power than $5\,$kW is provided and the probe's length is in the order of one meter.
	\item For very light melting probes without an additional actuator system, or if melting on icy bodies that are characterized by a low value of the gravitational acceleration, the melting head can become dangerously hot, and countermeasures have to be implemented.
\end{itemize}

Our model is applicable only to very simple probe geometries, i.e. probe's that have a flat melting head. In the future, it would be very interesting to also investigate the performance of  parabolic melting heads, as it is not obvious, how changing the head geometry interferes with the efficiency results we find in our paper. Though we believe that the adaption of our model to curved melting head geometries should be straight forward this remains to be future work. Another next step of our work will be a rigorous assessment of the initial phase of the melting mission, in which low pressure effects have to be taken into account and further complicate the physical process close to the surface.

\section*{Acknowledgement}
The project is supported by the Federal Ministry for Economic Affairs and Energy, Germany, on the basis of a decision by the German Bundestag (50 NA 1502). It is part of the Enceladus Explorer initiative of the DLR Space Administration.

\label{}



%\section*{References}
%\bibliographystyle{elsarticle-harv} 
\bibliographystyle{plainnat}
\bibliography{refs}


%\begin{thebibliography}{00}


%\bibitem{}

%\end{thebibliography}
\end{document}